\documentstyle[preprint,aps,epsf]{revtex}
\begin{document}
\draft
\title{\bf Analysis of quantum conductance 
 of  carbon nanotube junctions by the effective mass approximation}
\author{Ryo Tamura and Masaru Tsukada }
\address{Department of Physics, Graduate School of Science, University
  of Tokyo, Hongo 7-3-1, Bunkyo-ku, Tokyo 113, Japan}
\maketitle
\begin{abstract}
The electron transport through the nanotube junctions which
 connect the different metallic nanotubes by a pair of a pentagonal defect
 and a heptagonal defect is investigated by Landauer's formula
 and the effective mass approximation.
From our previous calculations based on the tight binding model, 
it has been known that the conductance is determined almost
 only by two parameters,i.e., the energy in the unit of the onset energy of more than two channels and the ratio of the radii of the two nanotubes.
The conductance is calculated again by the effective mass theory
in this paper and a simple analytical form of the conductance
 is obtained considering a special boundary conditions of 
the envelop wavefunctions.
The two scaling parameters appear naturally in this treatment.
The results by this formula coincide fairly well
 with those of the tight binding model.
 The physical origin of the scaling law is clarified by this
 approach.

\end{abstract}
\pacs{72.80.Rj,73.20.Dx,72.10.-d}
\narrowtext
Since the discovery of the carbon nanotubes \cite{tube}, 
they attract much interest
as one dimensional conductor with nanometer size.
It has been theoretically predicted that they become metallic or 
semiconducting according to the radius and the helicity 
of the honeycomb lattice forming  the tubes.\cite{tubetheory}
Electronic devices with nanometer size 
might be available by using the metallic nanotubes 
as leads.
Therefore we concentrate our discussion to the metallic nanotubes
in this paper.
Recently, it becomes possible to measure conductance of 
the individual nanotubes. \cite{experiment}
It is expected that such technological  progress
leads the realization of the nanometer devices.

The nanotubes can be combined naturally by introducing a pair of
a pentagonal defect and a heptagonal defect as observed by Iijima.
\cite{iijimajunction,tamurajunction1,tamurajunction2,saitojunction,chico}
Such defects are called disclinations and are necessitated to form
various  structures composed of curved surface of graphitic layer,
 for example, fullerenes,\cite{c60} minimal-surface structures \cite{minimal},
torus structures\cite{ihara}, cap structures at the end of the tube \cite{cap}
 and helical nanotubes \cite{akagi}.
The disclinations and their configurations have much influence in
both of their geometrical structures and electronic structures. \cite{discl}
Various remarkable examples have been reported on the crucial role of the disclination.
However, it is still unclear that what are the essential physics governing their effects on the electronic states of nano-structures.

In the preceding papers, 
we calculated the conductances of the junctions between the
two metallic nanotubes by using a tight
binding model including only the $\pi$ orbitals.\cite{tamurajunction1,tamurajunction2}
Our calculations based on Landauer's formula
have revealed a remarkable scaling law; 
the conductances of the junctions are 
determined almost only by two parameters in a certain Fermi
energy region.\cite{tamurajunction1}
Especially just at Fermi level of the undoped system, the conductance is almost
 determined only by the ratio of the circumference of the nanotubes.\cite{tamurajunction2}
The result indicates
that only the size and shape of the junctions are necessary to determine
 their conductances.Therefore the helicity of the
 honeycomb lattice is not relevant to determine the conductance, so far as 
 the junctions of the metallic nanotubes  are concerned.
It suggests that some continuum theory ignoring the atomic details but
 including only the size and shape of the junctions 
is valid, which would give some transparent view point of the scaling law.

Very recently, Matsumura and Ando proposed the effective mass approximations
 as  one of such continuum theories.\cite{matsumura}
They applied it to the electron transport of the nanotube junctions
 and made the physical meaning of the scaling law
somewhat clear. But their discussion is limited
just to the case of  the fixed Fermi energy case 
 of the undoped system and the conductance is not given 
as an analytical form, as far as we know.
In this paper, their discussion is generalized 
to the systems with general Fermi levels. 
Furthermore, the final result of
the conductance is represented by a closed analytical form
which gives a transparent view points.

First of all, we explain  the  Bloch state of  a monolayer graphite
 forming the single wall nanotube
by a simple tight binding model
 and relate it to the effective mass theory.\cite{ajiki}
Fig.\ref{tubetenkai}  shows the development map of the nanotube.
The vector $\vec{R}$ represents the circumference of the tube.
Two  parallel lines  perpendicular to $\vec{R}$ and parallel
 to the tube axis are connected with each other to form the tube.
Here we use two pairs of the 
 basic translation vectors $\{ \vec{e}_1,\vec{e}_2 \}$
and $\{ \vec{e}_{x'}, \vec{e}_{y'} \}$, where $\vec{e}_{x'}=(\vec{e}_1+\vec{e}_2)/\sqrt{3}$ and $\vec{e}_{y'}=\vec{e}_2-\vec{e}_1$, to represent components of vectors on the development map.
 For example, the components of $\vec{R}$  in Fig.\ref{tubetenkai} are represented as  $(R_1,R_2)=(2,5)$ and $(R_{x'},R_{y'})= (7\sqrt{3}/2,1.5)$.
In this paper, we concentrate our discussion to the metallic nanotube, 
 so that only the tube of which $R_1-R_2$ is an integer  multiple of three is considered.
\cite{tubetheory}
The four basic translation vectors
have the same length which is about 0.25 nm and
 denoted by $a$ hereafter.
The wavefunction can be represented by the Bloch state as $\psi_A(\vec{q})=
\exp (i(k_1 q_1+k_2 q_2))\psi_A(0)$ for sublattice $A$ and 
$\psi_B(\vec{q}+\vec{\tau})=\exp (i(k_1 q_1+k_2 q_2))\psi_B(\vec{\tau})$ for sublattice $B$, where $\vec{\tau}=(\vec{e}_1+\vec{e}_2)/3$,
$\vec{q}$ is the position of the atom with integer components
$q_1$ and $q_2$.
When the metallic nanotube is not doped,i.e.,the $\pi$ band is half filled,
 the Fermi energy locates 
at the $K$ and $K'$ corner points in the 2D Brillouin zone:
corresponding wavenumbers $(k_1,k_2)$ are $(2\pi/3,-2\pi/3)$ and
$(-2\pi/3,2\pi/3)$, respectively.
The corresponding energy  position, i.e., the Fermi level of the undoped
 system, is taken to be zero hereafter.
When the wavenumber $\vec{k}$ is near the  corner
point K, the wavenumber $\vec{k'}$ measured from the $K$ point,
$(k'_1,k'_2)=(-2\pi/3+k_1,2\pi/3+k_2)$, is small so that
the phase factors can be linearized as
 $\exp (ik_1a)=w \exp (ik'_1a) \simeq w(1+ik'_1a)$ and  $\exp (ik_2a)=w^{-1}
\exp (ik'_2a) \simeq w^{-1}(1+ik'_2a)$ , where $w=\exp (i2\pi/3)$.
Then  Schr$\ddot{\rm{o}}$dinger equation of a simple 
tight binding model for the Bloch state becomes
\begin{equation}
E \psi_A(\vec{q})=-i \gamma (wk'_2+w^{-1}k'_1)a\psi_B(\vec{q}+\vec{\tau})
\end{equation}
and
\begin{equation}
E \psi_B(\vec{q}+\vec{\tau})= i \gamma (w^{-1}k'_2+w k'_1)a\psi_A(\vec{q})\;\;.
\end{equation}
Here $\gamma$ is the hopping integral between the nearest neighboring sites, which is about $-3$ eV.
In this tight binding model, only the $\pi$ orbital is considered and mixing between the $\sigma$ and  the $\pi$ orbital caused by the finite curvature is neglected.
By using $\psi'_B(\vec{q})=(-i)\psi_B(\vec{q}+\vec{\tau})$
instead of $\psi_B$, one can obtain
\begin{equation}
E \psi_A(\vec{q})=\frac{\sqrt{3}}{2}\gamma a(k'_{x'}-ik'_{y'})\psi'_B(\vec{q})\;\;,
\label{blochA}
\end{equation}
and
\begin{equation}
E \psi'_B(\vec{q})=\frac{\sqrt{3}}{2}\gamma a(k'_{x'}+ik'_{y'})\psi_A(\vec{q})\;\;,
\label{blochB}
\end{equation}
where $k'_{x'}=(k'_1+k'_2)/\sqrt{3}$ and $k'_{y'}=k'_2-k'_1$.
The solution of these equations shows the linear dispersion relation,
\begin{equation} |\vec{k'}|=\pm \frac{2E}{\sqrt{3}\gamma a}\;\;.
\label{dispersion}
\end{equation}
For the one dimensional band which intersects the $K$ point,
the periodic boundary condition around the circumference is
$R_1 k'_1+R_2 k'_2=0$.
From this condition, one can show that phase difference 
between $A$ sublattice and $B$ sublattice is represented by
\begin{equation}
\psi'_B(\vec{q})/\psi_A(\vec{q})=\pm i  \exp (i\theta)\;\;,
\label{abphase}
\end{equation}
where $\theta$ is the angle of $\vec{R}$ with respect to $\vec{e}_{x'}$ measured  anti-clockwise as shown Fig.\ref{tubetenkai} .

In the effective mass theory, the wavefunction is represented  by
 $F_{A,B}^{K}(\vec{q})w^{(q_1-q_2)}$, where $F_{A,B}^{K}$ is the envelop wavefunction  and $w^{(q_1-q_2)}$ is the Bloch state wavefunction at the $K$ point.
When the Fermi level $E_F$ is close to zero,
the differential equation for $F_{A,B}^K$ is obtained from
eq.(\ref{blochA}) and eq.(\ref{blochB})  along a parallel discussion
 by replacement $k'_{x'} \rightarrow -i \partial_{x'}$ and $k'_{y'} \rightarrow -i \partial_{y'}$ as below.
\begin{equation}
(-\partial_{y'}-i \partial_{x'})F^K_B= k F^K_A
\end{equation}
\begin{equation}
(\partial_{y'}-i \partial_{x'})F^K_A= k F^K_B
\end{equation}
Here $k=|\vec{k'}|$ and the isotropic linear dispersion relation
(\ref{dispersion}) is used.
When $\tilde{F}_B^K= \exp (-i \theta) F_B^K$ is used instead of $F_B^K$,
 the above equations are transformed into 
\begin{equation}
(-\partial_{y}-i \partial_{x})\tilde{F}^K_B= k F^K_A\;\;,
\label{KA}
\end{equation}
and
\begin{equation}
(\partial_{y}-i \partial_{x})F^K_A= k \tilde{F}^K_B\;\;,
\label{KB}
\end{equation}
where $(x,y)$ is defined by
\begin{equation}
\partial_{y}-i \partial_{x} =(\partial_{y'}-i \partial_{x'})\exp (-i \theta)\;\;.
\label{rotate}
\end{equation}
The above transformation corresponds to the rotation of coordinate axes
from $\{\vec{e}_{x'},\vec{e}_{y'} \}$ to $\{\vec{e}_{x},\vec{e}_{y} \}$
by the angle $\theta$ anti-clockwise so that the $x$ axis is parallel to
the circumference vector $\vec{R}$ as shown in Fig.\ref{tubetenkai}.
In the following discussion, the $(x,y)$ coordinates are used and
$\tilde{F}_B^K$ is written simply as $F_B^K$ for simplicity.
It can be seen from eq.(\ref{abphase}) that the Bloch state wavefunction
 for the one dimensional band intersecting the $K$ point is represented by $(F_A^K, F_B^K)=(1, \pm i)$.
The equations of the envelop wavefunctions $F_{A,B}^{K'}$ for
 the $K'$ corner point can be easily obtained in a similar way
 as
\begin{equation}
(\partial_{y}-i \partial_{x})F^{K'}_B= k F^{K'}_A\;\;,
\label{K'A}
\end{equation}
and
\begin{equation}
(-\partial_{y}-i \partial_{x})F^{K'}_A= k F^{K'}_B\;\;.
\label{K'B}
\end{equation}
Hereafter the envelop wavefunctions $F$'s 
are simply called the wavefunctions.

The wavefunction in the nanotube is given by the plane wave;
$\exp (i(k_x x \pm  k_y y))$. When the $x$ direction is taken to be parallel
to the circumference of the tube, $k_x$ is quantized as
$k_x(n)=2\pi n/R$ and $k_y$ is given by $k_y(n)=\sqrt{k^2-k_x(n)^2}$.
Here $n$ is an integer representing a number of nodes around the
 circumference and $R$ is the circumference of the tube.
When $k_y(n)$ is a real number, the channel $n$ is open and
 the corresponding wavefunction is extended, 
 otherwise the channel is closed and the wavefunction shows exponential
 grow or decay.
The number of the open channel is called the channel number.
When the Fermi energy is zero,
 only the channel $n=0$ is open, and therefore the channel number is 
 kept  two irrespective of $R$.

The electronic states near the  Fermi energy for the undoped system
($E_F=0$) govern the electron transport,
so discussion in this paper is concentrated to  this Fermi energy region
 where the channel number is kept two.
In order to discuss the wavefunction in the junction
part, the polar coordinates $(r,\theta)$ is useful.
Its relation to the coordinate $(x,y)$ is 
usual one,i.e.,$ r=\sqrt{x^2+y^2}$, $\tan \theta =y/x$.
Fig.\ref{junctiontenkai} is the development map of the nanotube junction where
 the coordinate $(x,y)$ is defined.\cite{saitojunction}
A heptagonal defect and a pentagonal defect are introduced
at A(=B) and C(=D), respectively.
 Therefore the indexes '7' and '5' are
 used to represent the thinner tube and thicker tube, respectively.
The equilateral triangles '$\Delta$ OAB' and '$\Delta$ OCD' with bases 'AB' and 'CD' have common apex 'O',
 which is taken to be the origin of the coordinate
$(x,y)$ in this paper. 
Then the wavefunction satisfies the wave equation $(z^2\partial_z^2-z\partial_z-\partial_{\theta}^2-z^2) F = 0$, where $z=kr$.
The solution is represented by Bessel functions $J_m$ and Neumann functions
 $N_m$  as 
\begin{equation}
F= \sum_{m=-\infty}^{\infty} e^{im\theta}(c_m J_{|m|}(z)+d_m N_{|m|}(z)) \;\;.
\label{JmNm}
\end{equation}

Matsumura and Ando have found that the wavefunction should satisfy
the following boundary conditions in the junction part.\cite{matsumura}
\begin{equation}
F^{K'}_A(z,\theta+\pi/3)=-F^K_B(z,\theta)
\label{bound1}
\end{equation}
\begin{equation}
F^{K}_B(z,\theta+\pi/3)=-\frac{1}{w}F^{K'}_A(z,\theta)
\label{bound2}
\end{equation}

\begin{equation}
F^{K}_A(z,\theta+\pi/3)=F^{K'}_B(z,\theta)
\label{bound3}
\end{equation}
\begin{equation}
F^{K'}_B(z,\theta+\pi/3)=wF^{K}_A(z,\theta) 
\label{bound4}
\end{equation}
In these equations, $w \equiv e^{i2\pi/3}$.
From eq.(\ref{bound1}) and eq.(\ref{bound2}), terms in eq.(\ref{JmNm})
for $F^{K'}_A$ and $F^K_B$ are not zero only when $m=3 l+2 \;\;(l=$integer
).
 Because the open channel $n=0$ has no node along the circumference,
 it is fitted to the components with smaller $|m|$ in eq.(\ref{JmNm}) 
better than to those with larger $|m|$.
So we assume that one can neglect all the terms except those with $l=0$ and $l=-1$ in eq.(\ref{JmNm}). Then the wavefunctions can be written as
\begin{equation}
F^{K'}_A=e^{2i\theta} f_2(z)+e^{-i\theta} f_1(z),
\end{equation}

and

\begin{equation}
F^{K}_B=-e^{2i(\theta+\frac{\pi}{3})} f_2(z)-e^{-i(\theta+\frac{\pi}{3})} f_1(z)\;\;,
\end{equation}

where

\begin{equation}
f_m(z)=c_mJ_m(z)+d_mN_m(z) \;\;(m=1,2).
\end{equation}

From eq.(\ref{KA}) and eq.(\ref{K'B}), the other two wavefunction $F^{K'}_B$ and $F^K_A$ can be derived from $F^{K'}_A$ and $F^K_B$ as

\begin{equation}
F^{K'}_B=-i e^{i\theta} \tilde{f}_2(z)+ie^{-i2\theta} \tilde{f}_1(z) \;\;,
\label{F4}
\end{equation}

\begin{equation}
F^{K}_A=i e^{i(\theta+\frac{2\pi}{3})} \tilde{f}_2(z)-ie
^{-i(2\theta+\frac{\pi}{3})} \tilde{f}_1(z) \;\;,
\label{F1}
\end{equation}

where

\begin{eqnarray}
\tilde{f}_1(z)=c_1J_2(z)+d_1N_2(z),\nonumber \\
\tilde{f}_2(z)=c_2J_1(z)+d_2N_1(z)\;\;,
\end{eqnarray}
by using well known recursion formula of the Bessel functions and Neumann
 functions.
It is easily confirmed that eq.(\ref{F4}) and eq.(\ref{F1}) satisfy
 the boundary conditions eq.(\ref{bound3}) and eq.(\ref{bound4}).
The amplitude of the open channel in the tube, which is denoted by $\alpha$,  is obtained from the wavefunctions in the junction part as
\begin{equation}
\alpha_{7\pm}^K=\frac{1}{\sqrt{2R_7}}\int_A^B  dx (F^K_A \pm i F_B^K)\;\;,
\label{a7}
\end{equation}

and

\begin{equation}
\alpha_{5\pm}^K=\frac{1}{\sqrt{2R_5}}\int_C^D  dx' (F^K_A \pm i\exp (i \phi) F_B^K)\;\;,
\label{a5}
\end{equation}
for the $K$ point.
The indexes $+$ and $-$ mean directions along which the electronic waves
propagate as shown in Fig.\ref{junctiontenkai}. $R_5$ and $R_7$ are the circumferences of
 the thicker tube and thinner tube, and $\phi$ is angle
 between 'AB' and 'CD' in the development map Fig.\ref{junctiontenkai}.\cite{saitojunction}
Equations for the $K'$ points  are obtained from eq.(\ref{a7}) and eq.(\ref{a5}) by replacing $i$  and $\phi$ 
with $-i$ and $-\phi$, respectively.
To simplify the calculation, the integrations
in the above equations are transformed as
$\int_C^D  dx' \rightarrow R_5 \int_{-\frac{2}{3}\pi-\phi}
^ {-\frac{\pi}{3}-\phi}  d\theta$ and
$\int_A^B  dx \rightarrow R_7 \int_{-\frac{2}{3}\pi}
^ {-\frac{\pi}{3}} d\theta$.
If the wavefunction in the junction varies slowly  near $r=R_5$ and $r=R_7$
 along the radial directions,
 this replacement can be allowed.

The relation among the amplitudes of the open channel in the thicker tube,
$\vec{\alpha}_5=\; ^t\,(\alpha_{5+}^K,\alpha_{5+}^{K'},\alpha_{5-}^K,
\alpha_{5-}^{K'} )$, those in the thinner tube,
$\vec{\alpha}_7=\; ^t\,(\alpha_{7+}^K,\alpha_{7+}^{K'},\alpha_{7-}^K,
\alpha_{7-}^{K'} )$ and the wavefunctions in the junction part,
 $\vec{c}=\; ^t\,(c_2,d_2,c_1,d_1 )$ is summarized in the followings.

\begin{equation}
\vec{\alpha}_5=\sqrt{R_5}P_5(\phi) M \Lambda(\phi) L(k R_5 ) \vec{c}
\label{a5mat}
\end{equation}

\begin{equation}
\vec{\alpha}_7=\sqrt{R_7}P_7 M L(k R_7 ) \vec{c}\;\;,
\label{a7mat}
\end{equation}
where $M$ are constant matrixes as
\begin{equation}
M=
\left( \begin{array}{cccc}
w & 0 & 0 & -\frac{\sqrt{3}}{2}iw \\
0 & \frac{\sqrt{3}}{2}w & iw & 0 \\
0 &  -\frac{\sqrt{3}}{2}& i & 0 \\
-1 & 0 & 0 &  -\frac{\sqrt{3}}{2}i 
\end{array} \right)\;\;.
\end{equation}
$\Lambda(\phi)$ is a diagonal matrix, where $\Lambda_{1,1}=\Lambda_{3,3}^*=\exp (-i \phi)$ and $\Lambda_{2,2}=\Lambda_{4,4}^*=\exp (-2i \phi)$.
$P_7$ and $P_5 (\phi)$ are defined by eq.(\ref{a7}) and eq.(\ref{a5}), respectively. The matrix elements of  $L(z)$ are $L_{11}=L_{33}=J_1(z)$,$L_{12}=L_{34}=
N_1(z)$,$L_{21}=L_{43}=J_2(z)$ and $L_{22}=L_{44}=N_2(z)$. The other matrix
 elements of $L(z)$ are zero.
The vector $\vec{\alpha}_7$ is given by a linear transformation of 
$\vec{\alpha}_5$ defined by eq.(\ref{a5mat}) and eq.(\ref{a7mat}).
The transformation matrix $T$ (i.e., $\vec{\alpha}_7=T\vec{\alpha}_5$)
 includes three independent parameters, i.e.,
 $\beta \equiv R_7/R_5$, $z_5 \equiv k R_5$ and $\phi$.
The parameter $z_5$ is related to the Fermi Energy $E_F$ as follows.
When  $|k|$ is near zero, channel number is always two
which is independent of the radius of the nanotubes.
But as $|k|$ increases, the channel number 
increases firstly in the thicker tube, when $|k|$ exceeds $k_c=2\pi/R_5$ .
Owing to the linear dispersion relation eq.(\ref{dispersion}), 
$z_5=2 \pi k/k_c= 2 \pi E_F/E_c$ holds where
$E_c$ is the threshold Fermi energy corresponding to $k_c$.
The transmission rates are calculated from the transfer matrix $T$,
and the conductance $\sigma$ is obtained by Landauer's formula
 as 

\begin{equation}
\sigma=\frac{24}{
\{\beta\sum_{i=1}^2\sum_{j=1}^2 
(3/4)^{i+j-2}X_{i,j}^2 
 \} + 6 }
\label{result}
\end{equation}

where

\begin{eqnarray}
X_{i,j}=\pi z_5 \{J_i (\beta z_5) N_j(z_5)-N_i(\beta z_5) J_j
(z_5)\}\;\;.
\end{eqnarray}

The obtained conductance has a remarkable feature that it does not
 depend on the angle $\phi$.
It is consistent with the scaling law with the
 two parameters $E_F/E_c$ and $\beta=R_7/R_5$ in Ref.\cite{tamurajunction2}.
When $E_F=0$, $\sigma$ has a simpler form as
\begin{equation}
\sigma=8/(\beta^3 + \beta^{-3}+2)\;\;.
\label{zerosigma}
\end{equation}
Eq.(\ref{zerosigma}) is also fitted well to the numerical data in Ref.\cite{tamurajunction1} which shows one parameter scaling law.

Fig.\ref{negative} and Fig.\ref{positive} show the 
comparison between the conductances obtained by the tight binding model and those by eq.(\ref{result}) for
 the  $(2i,2i+3)$-(10,13) junctions, where the same notation
 as that of Ref.\cite{tamurajunction1}  is used to specify the junctions.
The horizontal axis is $|E_F|$ with the unit of the absolute value
 of the hopping integral $|\gamma|$.
The figures are shown in the energy range where the channel number 
 is kept two, i.e., $|E_F|<E_c$.
Agreement between the two is quite good.
But difference is larger for positive Fermi energy 
 than that for negative Fermi energies.
This difference is explained as the effect of discreteness of the lattice
as follows.
The wavefunction tends to have opposite signs between neighboring sites
 when the Fermi energy is positive.
But such alternating sign arrangements are not possible
for the paths with odd number steps
 surrounding the odd membered ring defects.
Such effects come from the discreteness of the lattice
 and can not be treated by the continuum model assumed 
 in the the effective mass theory.
It should be also noted that a drop of $\sigma$ near $E_F=E_c$
appears in the tight binding model,
while it is not reproduced by eq.(\ref{result}).
It comes from the effects of the closed channel which are not
included in eq.(\ref{result}).
As $E_F$ approaches $E_c$, the decay length
 of the closed channel $n=1$ increases and becomes more extended
 so that it has more important role in the transport. 
When the values of $\beta=R_7/R_5$ approaches to unity  1, the conductance 
 does not depend on $E_F$ taking the values close to the maximum value.
As the parameter $\beta$ decreases, the conductance $\sigma$ near $E_F=0$
decreases and peak structure appears at $E_F$ somewhat lower
than $E_c$.
When $\beta$ increases further, the height of the peak is lowered.
 The  value of $z_5$ corresponding to the Fermi energy
 at the peak is about half of the first period of $J_2(z)$,
 and 3/4 of that of $J_1(z)$ so that we speculate that the peak structures
are formed by resonance of these radial wavefunctions.
More detailed discussions will appear elsewhere.

In this paper, the scaling law of the conductance
of the nanotube junctions previously found by the tight binding model
 can be reproduced and analyzed by
 the effective mass theory. It provides the analytical representation 
of the conductances and a clear physical interpretations.
It can be also applied to various structures relating to the nanotubes.

We would like to thank H. Matsumura and T. Ando for their useful
 suggestions.
This work has been supported by the Core Research for Evolutional 
 Science and Technology (CREST) of the Japan Science and Technology 
 Corporation(JST).

\begin{figure}

\caption{Development map of the nanotube.}
\label{tubetenkai}

\end{figure}

\begin{figure}

\caption{Development map of the nanotube junctions.
It is similar to that of Ref. \protect\cite{saitojunction}.
Lines 'EB' and 'FA' are parallel.
Lines 'DG' and 'CH' are also parallel and their angle with respect to 
'EB' and 'FA' is denoted by $\phi$. The lines 'EB','BD' and 'DG'
 are connected with the lines 'FA','AC' and 'CH', respectively.
The rectangles 'EBAF' and 'DGHC' form the thinner tube and the thicker
 tube, respectively. 'BD' is the rotated 'AC' by angle of
 60 degree and the quadrilateral 'ABDC' forms a junction part with a shape
of a part of a cone. 
A heptagonal defect and a pentagonal defect are introduced
at A(=B) and C(=D), respectively.}
\label{junctiontenkai}

\end{figure}

\begin{figure}

\caption{The conductances of the junction between $(2i,2i+3)$ tube
and (10,13) tube  for negative Fermi energy.
Corresponding values of $i$
are attached to each graphs.
 The horizontal axes are the absolute value
of the Fermi energy in units of the absolute value of the hopping integral
$|\gamma|=-\gamma$. The hopping integral $\gamma$ is taken to be negative.
They are calculated by eq.(\protect\ref{result}), and the tight binding model,
 each of which are represented by dotted lines and plots connected with solid
 lines.}

\label{negative}
\end{figure}

\begin{figure}

\caption{The same figure as Fig.2 for the positive Fermi energy.}
\label{positive}
\end{figure}

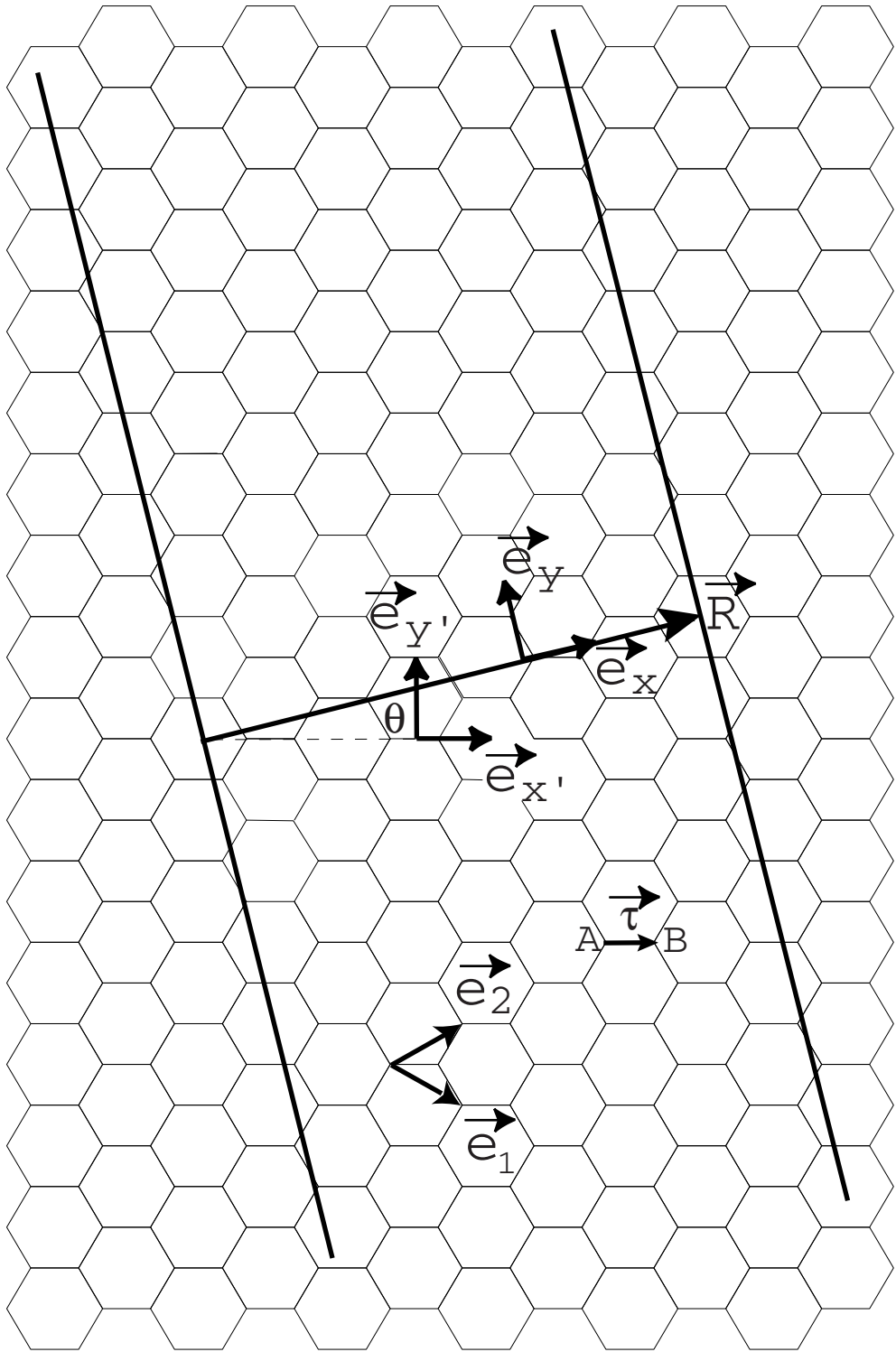
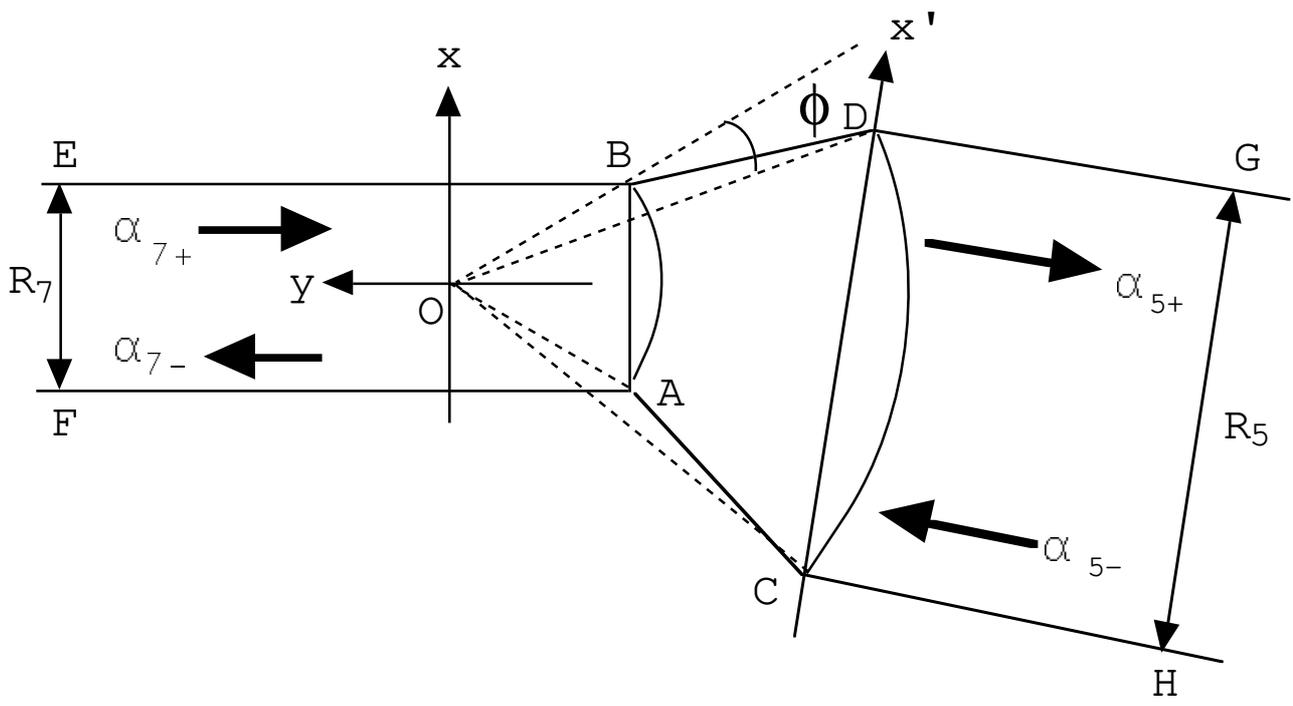
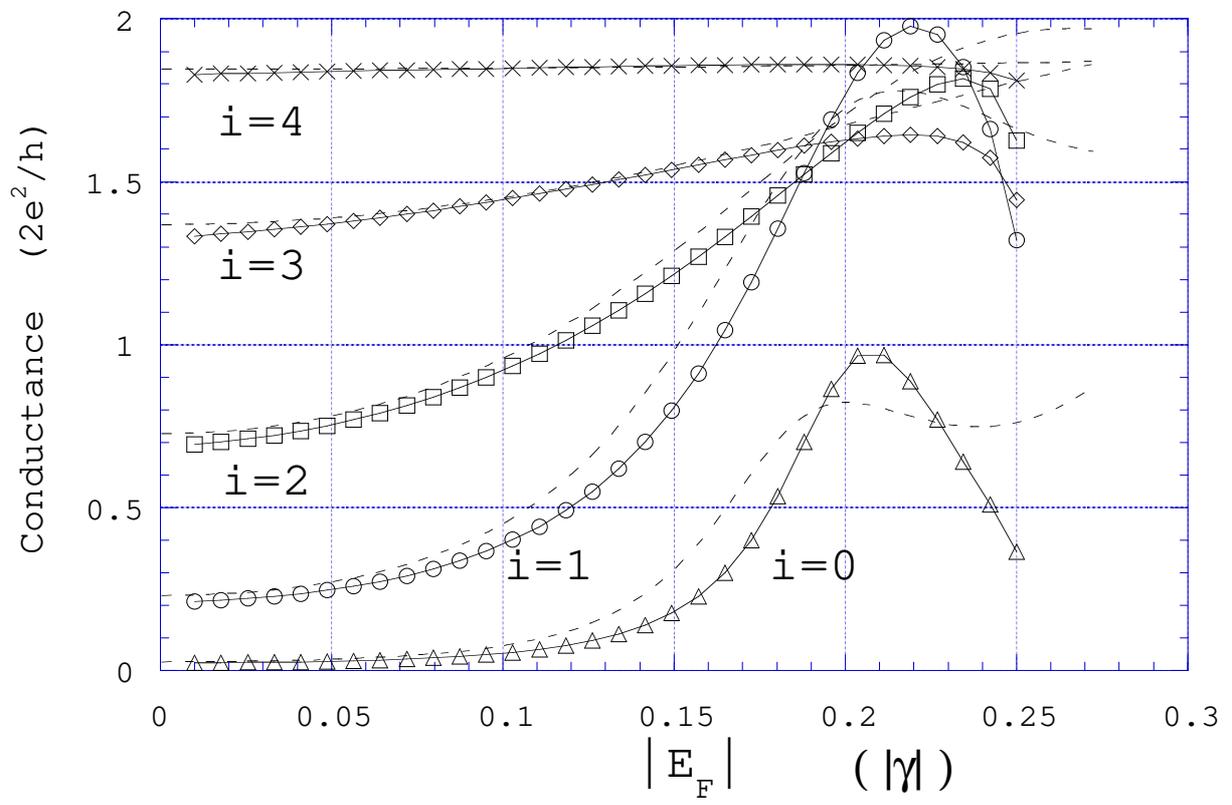
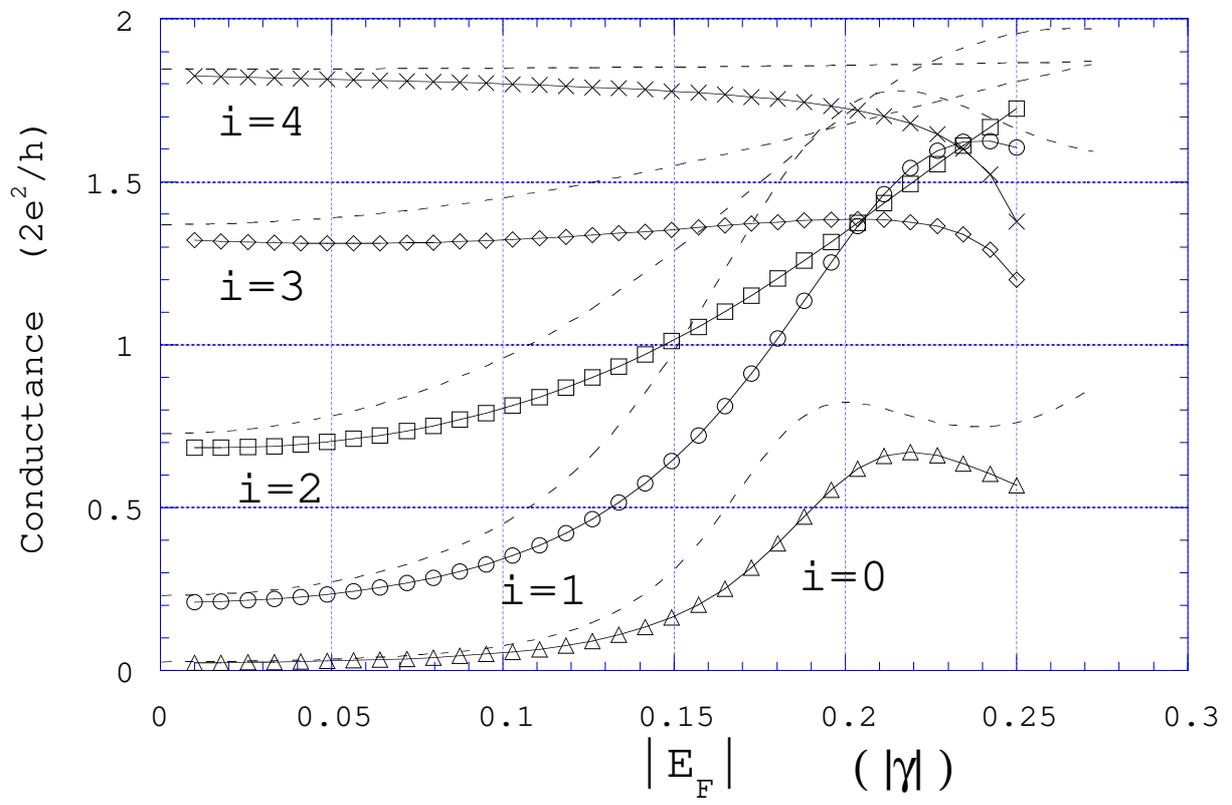
\end{document}